\documentclass[conference]{IEEEtran}[2015/08/26]

\usepackage{mwe}

\usepackage[T1]{fontenc}
\usepackage[utf8]{inputenc} 

\usepackage{graphicx}

\usepackage[ngerman,main=english]{babel}
\addto\extrasenglish{\languageshorthands{ngerman}\useshorthands{"}}

\usepackage{upquote}

\usepackage{paralist}

\usepackage{csquotes}

\usepackage{url}
\makeatletter
\g@addto@macro{\UrlBreaks}{\UrlOrds}
\makeatother

\usepackage{booktabs}

\usepackage{xcolor}

\usepackage{listings}
\usepackage{pdfcomment}
\ifCLASSOPTIONcompsoc
  \usepackage[%
    square,        
    comma,         
    numbers,       
    sort           
  ]{natbib}
\else
  \usepackage[%
    square,        
    comma,         
    numbers,       
    sort&compress 
  ]{natbib}
\fi

\usepackage{etoolbox}
\makeatletter
\patchcmd{\NAT@test}{\else \NAT@nm}{\else \NAT@hyper@{\NAT@nm}}{}{}
\makeatother

\usepackage{hyperref}
\hypersetup{hidelinks,
  colorlinks=true,
  allcolors=black,
  pdfstartview=Fit,
  breaklinks=true}
\usepackage[all]{hypcap}

\usepackage[capitalise,nameinlink]{cleveref}
\crefname{lstlisting}{\lstlistingname}{\lstlistingname}
\Crefname{lstlisting}{Listing}{Listings}

\newenvironment{listing}[1][htbp!]{\begin{figure}[#1]}{\end{figure}}

\usepackage{xspace}

\ifCLASSOPTIONcompsoc
  \usepackage[caption=false,font=footnotesize,labelfont=sf,textfont=sf]{subfig}
\else
  \usepackage[caption=false,font=footnotesize]{subfig}
\fi

\usepackage{stfloats}

\newcommand{\fio}{FACTORY~I/O}

\newcommand{\opc}{OpenPLC}

\newcommand{\mb}{Modbus}

\begin{document}

\title{Towards Flexible Security Testing of OT Devices}

\author{%
  \IEEEauthorblockN{Florian Wilkens}
  \IEEEauthorblockA{Universität Hamburg, Germany\\
    wilkens@informatik.uni-hamburg.de}
  \and
  \IEEEauthorblockN{Samuel Botzler}
  \IEEEauthorblockA{Universität Hamburg, Germany\\
    7botzler@informatik.uni-hamburg.de}
  \and%
  \IEEEauthorblockN{Julia Curts}
  \IEEEauthorblockA{Universität Hamburg, Germany\\
    7curts@informatik.uni-hamburg.de}
  \and%
  \IEEEauthorblockN{Skadi Dinter}
  \IEEEauthorblockA{Universität Hamburg, Germany\\
    6dinter@informatik.uni-hamburg.de}
  \and%
  \IEEEauthorblockN{Malte Hamann}
  \IEEEauthorblockA{Universität Hamburg, Germany\\
    mhamann@informatik.uni-hamburg.de}
  \and%
  \IEEEauthorblockN{Vincent Hubbe}
  \IEEEauthorblockA{Universität Hamburg, Germany\\
    7hubbe@informatik.uni-hamburg.de}
  \and%
  \IEEEauthorblockN{Aleksandra Kornivetc}
  \IEEEauthorblockA{Universität Hamburg, Germany\\
    6kornive@informatik.uni-hamburg.de}
  \and%
  \IEEEauthorblockN{Nurefsan Sertbas}
  \IEEEauthorblockA{Universität Hamburg, Germany\\
    sertbas@informatik.uni-hamburg.de}
  \and%
  \IEEEauthorblockN{Mathias Fischer}
  \IEEEauthorblockA{Universität Hamburg, Germany\\
    mfischer@informatik.uni-hamburg.de}
}


\maketitle

%
%

\begin{abstract} 
In the factory of the future traditional and formerly isolated Operational Technology (OT) hardware will become connected with all kinds of networks. This leads to more complex security challenges during design, deployment and use of industrial control systems. As it is infeasible to perform security tests on production hardware and it is expensive to build hardware setups dedicated to security testing, virtualised testbeds are gaining interest.
We create a testbed based on a virtualised factory which can be controlled by real and virtualised hardware. This allows for a flexible evaluation of security strategies.

\end{abstract}

\section{Introduction}\label{sec:introduction}
Current Industrial Control Systems (ICS) have been designed to monitor and control industrial processes in isolated OT networks. With the advent of Industry 4.0 and the Industrial Internet of Things (IIoT), these OT networks more and more converge with other networks and the Internet. This convergence increases functionality and effectiveness, while reducing costs. However, this brings new vulnerabilities and opens these networks to threats and attacks from the Internet. However, applying traditional IT security solutions alone will fail, as these solutions do not take into account the cyber and physicial components of an ICS \cite{weiss}.

It is important to understand the potential effects of attacks on critical infrastructure and how to defend them. For that, a security assessment of OT systems is required, in the design phase, but also when deployed. However, in most cases it is not possible to run security tests on production systems due to availability and performance concerns. For this reason, it is necessary to build an experimental platform to develop and evaluate cyber security solutions for OT environments. This allows to conduct security tests, so that there is no side effect on the production. However, constructing a testbed is challenging even with the help of modern technologies, as it can be difficult to obtain a realistic testbed scale and configuration \cite{holm2015survey}.

There are some criteria that needs to be considered in testbed development phase. First of all, it is significantly important that how accurately the testbed reproduces a real system. In many cases, reproducing every details of a system is not needed but the testbed should offer reasonable level of realism. In this way, researchers may evaluate their hypothesis on the testbed. In addition to that, experiments should be able to produce statistically consistent results for the repetitions. Also, testbed should provide isolation for the experiments so that external activities do not affect the results \cite{siaterlis2012use}.

In this paper, we present our OT security testbed based on \fio\footnote[2]{\url{https://factoryio.com/}} and \opc\footnote[3]{\url{https://www.openplcproject.com/}}. The proposed testbed provides a platform to implement and test security mechanisms in a wide range of use cases, e.g., simulating complex attack scenarios or creating a dataset by generating traffic from a large number of devices. It allows to evaluate security mechanisms and thus eases the research on OT security.

The remainder of this paper is structured as follows: Section \ref{sec:relatedwork} gives an overview on related work in the area of ICS security. The design goals and architecture of the testbed are explained in Section \ref{sec:architecture}. The real-world implementation, including setup, possible attacks and monitoring approaches is described in Section \ref{sec:evaluation}. We conclude the paper in Section \ref{sec:conclusion}.

\section{Related Work}\label{sec:relatedwork}
There have been several attempts to develop research testbeds for industrial control systems. These testbed constructions can be categorised into four groups based on the used technologies as physical replication, simulation and emulation, virtualisation, and hybrid testbeds.

To study the vulnerability of ICS systems, researchers attempt to replicate the existing system in isolated environments named \textbf{physical replication} testbeds \cite{holm2015survey}. In \cite{green2017pains}, a physical ICS testbed is created and some attacks are performed to show how the attacker may gain information from the environment. However, the proposed testbed is not easily reproducible and would require purchasing a lot of hardware.  A power-generation testbed consisting of Human-Machine-Interfaces (HMIs), Programmable Logic Controllers (PLCs), and AC and DC motor pairs has been developed for simulating the electricity generation. In these testbeds, carrying out security related tests and evaluating the vulnerabilities provides a  high degree of accuracy. However, it is difficult to reconfigure and maintain real hardware and software in a testbed \cite{qassim2017survey}.

The second method to build a testbed is \textbf{simulating and emulating} system components. For instance, 
network simulation tools like ns2, OPNET, OMNet++, SSFnet, RINSE can be used for simulating and emulating control traffic networks. To simulate the communication of PLCs and other field devices, tools like STEP7, RSEmulate, Modbus Rsim, Soft-PLC are suitable. In order to model the physical processes, Matlab, Modelica, Ptolemy, PowerWorld are possisble tools \cite{geng2019survey}.
Software based simulation is an easy and reusable way of building testbeds. It enables deploying a rich set of scenarios for security testing and evaluation. However, it does not reflect the real vulnerabilities of the systems due to being limited to how it is coded. 

\textbf{Virtualisation} is not clearly defined in the area of ICS research and often there is a smooth transition between simulation, emulation, and virtualisation. In most cases a testbed is called virtual or virtualised if it contains some or all components built in software. Virtualisation techniques, e.g. VirtualBox, GENI, and PlanetLab, are mostly used for virtualising control centers and communication components \cite{holm2015survey}. In general, virtualisation is a good approach for obtaining a large-scale realistic testbed. Carrying out security related tests on a real automation hardware may be more realistic, but also more expensive and less flexible. Thus, deploying environment-specific, virtualised testbeds is getting interest~\cite{holm2015survey}.

\textbf{Hybrid} testbeds combine software and hardware components. In \cite{xie2018vtet}, the authors simulate a chemical process in a testbed called VTET that contains software and hardware components. They also describe different attack scenarios which are targeting the process to damage the physical devices. Another hybrid testbed for studying cyberattacks was introduced in \cite{reaves2012open}. It contains a lot of software components written in python that can be combined with hardware communication devices and are interoperable with control hardware.

\section{Testbed Architecture\label{sec:architecture}}
The main design goal for the testbed is a flexible architecture that allows to combine multiple hardware and software components while keeping the costs low. A managed Ethernet switch provides the necessary configurable connectivity for three types of components in the testbed: 1. factory entities, 2. controlling entities, 3. security entities. 

\begin{figure}
    \centering
    \includegraphics[width=.9\linewidth]{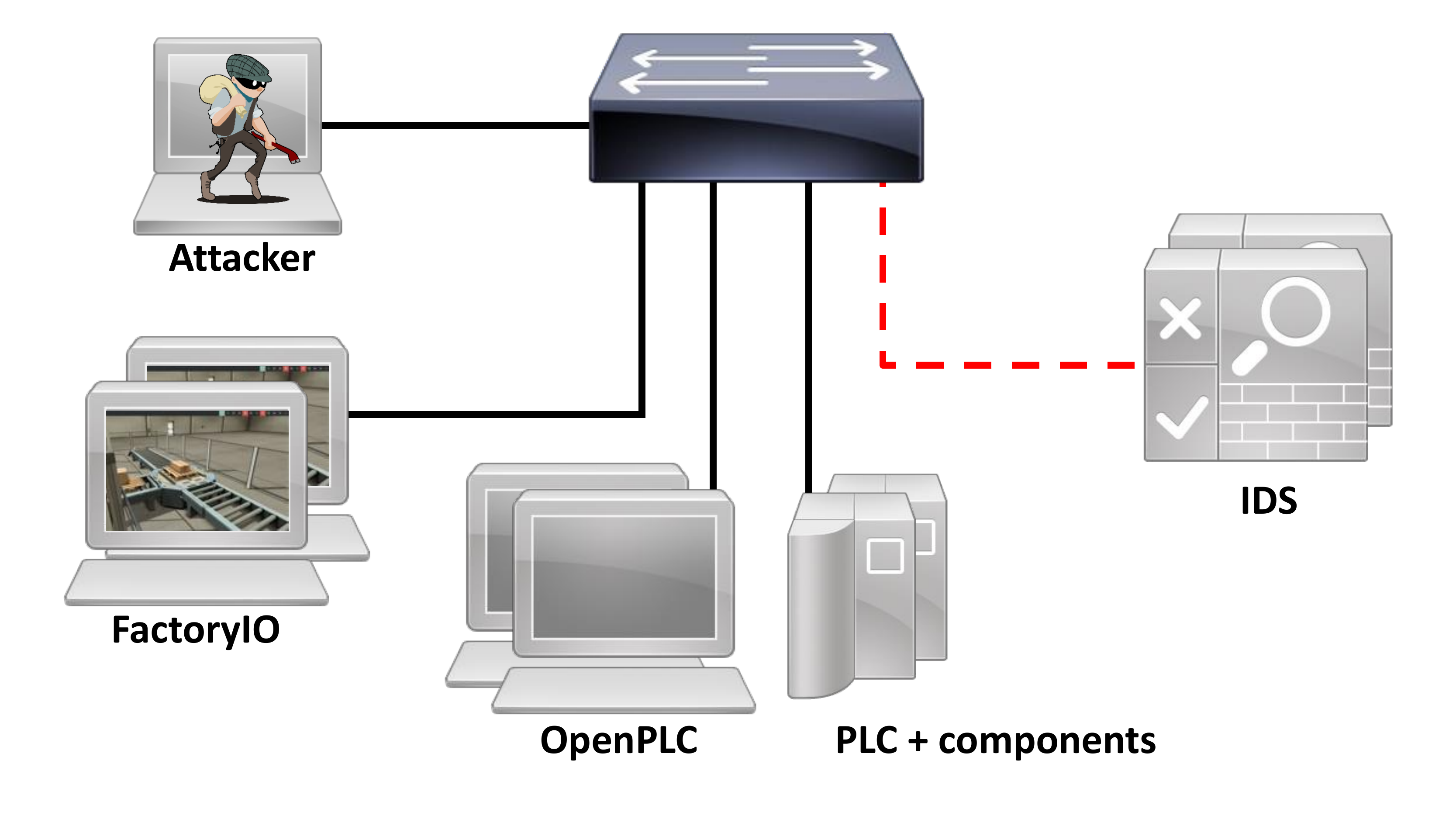}
    \caption{\footnotesize Testbed Overview. Hosts running \fio\ are directly connected to the managed switch. Optionally physical PLCs and/or hosts running OpenPLC are also connected on regular ports (black/solid), while the IDS receives all produced traffic via a mirror port (red/dashed).\label{fig:overview}}
\end{figure}

A generic overview of the testbed is given in Figure~\ref{fig:overview}. A small factory can be built in \fio\ which simulates and renders an industry process, e.g., a conveyor belt and a robot sorting the items arriving on the conveyor belt. Hardware components of a factory like an HMI can be integrated via a PLC or connected via ethernet if possible. As \fio\ offers a lot of different drivers for external controllers it can be steered by Siemens PLC, Allen-Bradley PLC, \mb, OPC-UA and others. This allows to add controlling entities as hardware (PLC) and as software, with a lot of possible host devices. To get the contol data sent over the network to the monitoring and security entities, the switch mirrors all traffic and forwards it them. A security entity can be any kind of machine that runs some monitoring tool, software, or an Intrusion Detection System (IDS). As it can be replaced fexibly, we can employ several monitoring tools, anomaly detection tools and even machine learning algorithms.

By connecting another host the switch an attacker that already gained access to the network can be simulated. There are multiple possible attacks in this scenario. The attacker can monitor the traffic, learn communication patterns, disrupt communications (DoS), and forge malicious control messages. These can then disrupt the production, destroy equipment or even cause harm to workers. Additional threats are espionage and manipulations which increase the scrap rate gradually over time while keeping a low profile.


\section{Testbed Implementation\label{sec:evaluation}}

We built a testbed according to the architecture described in Section~\ref{sec:architecture}. Additionally, we focused on ease of demonstration to highlight the impact of attacks to people unfamiliar with OT.
In the following, we describe our testbed in detail, show how we successfully attacked the simulated factory and outline monitoring solutions.

\subsection{Setup\label{subsec:setup}}
\begin{figure}
    \centering
    \includegraphics[width=.9\linewidth]{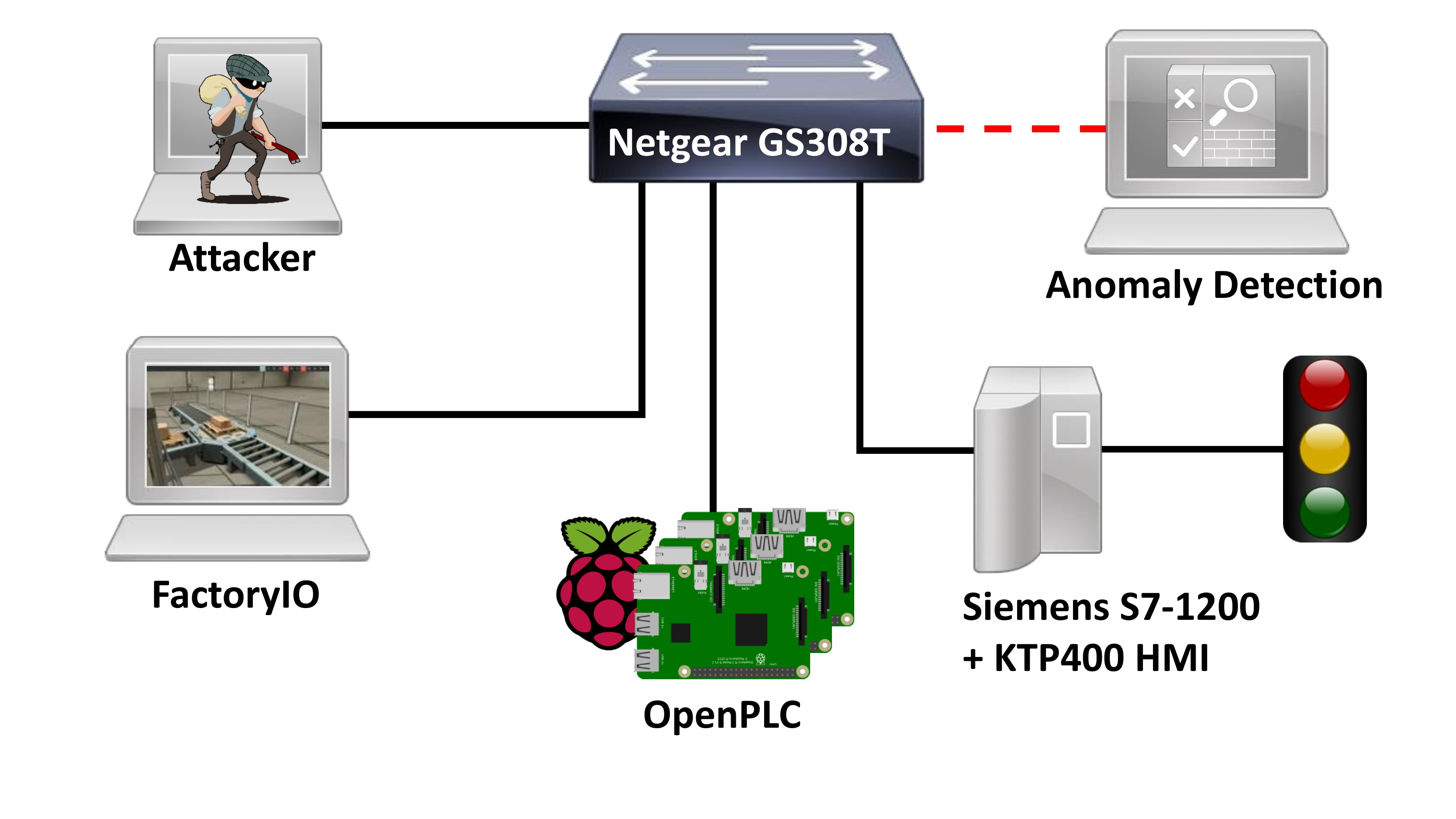}
    \caption{Hardware overview\label{fig:hardware-overview}}
\end{figure}

\autoref{fig:hardware-overview} gives an overview on the hardware components in our OT testbed. Our virtual factory in \fio\ runs on Windows on a regular Dell workstation. Another identical machine is used to run monitoring software, i.e., an IDS as well as the anomaly detection scripts that we describe later on. The automation workload is executed on one physical PLC, a Siemens S7--1200 PLC, and three RasperryPis running \opc\ on Raspbian Linux. Additionally we added a Siemens KTP400 HMI and a three-colored warning light. All networked devices are connected via a managed switch from Netgear with a mirror port the IDS is connected to. During the attack simulation later on we added a Linux machine also connected to a regular port to match the conditions an attacker would likely face.


We chose commodity hardware for both the Windows computers as well as the machines running \opc\ and the switch. The Siemens PLC and HMI were integrated to offer a physical feedback link. We feel, that this mixture between real automation hardware and commodity machines provides the best trade-off between minimizing costs and staying close to real-world deployments. Additionally, the physical devices such as the warning light help immensely when demoing the testbed to people who are not completely familiar with the OT world. For example, a red warning light that flashes after an emergency button was pushed on the HMI provides instant feedback and helps the viewer to understand what is happening.

\begin{figure}
    \centering
    \includegraphics[width=.9\linewidth]{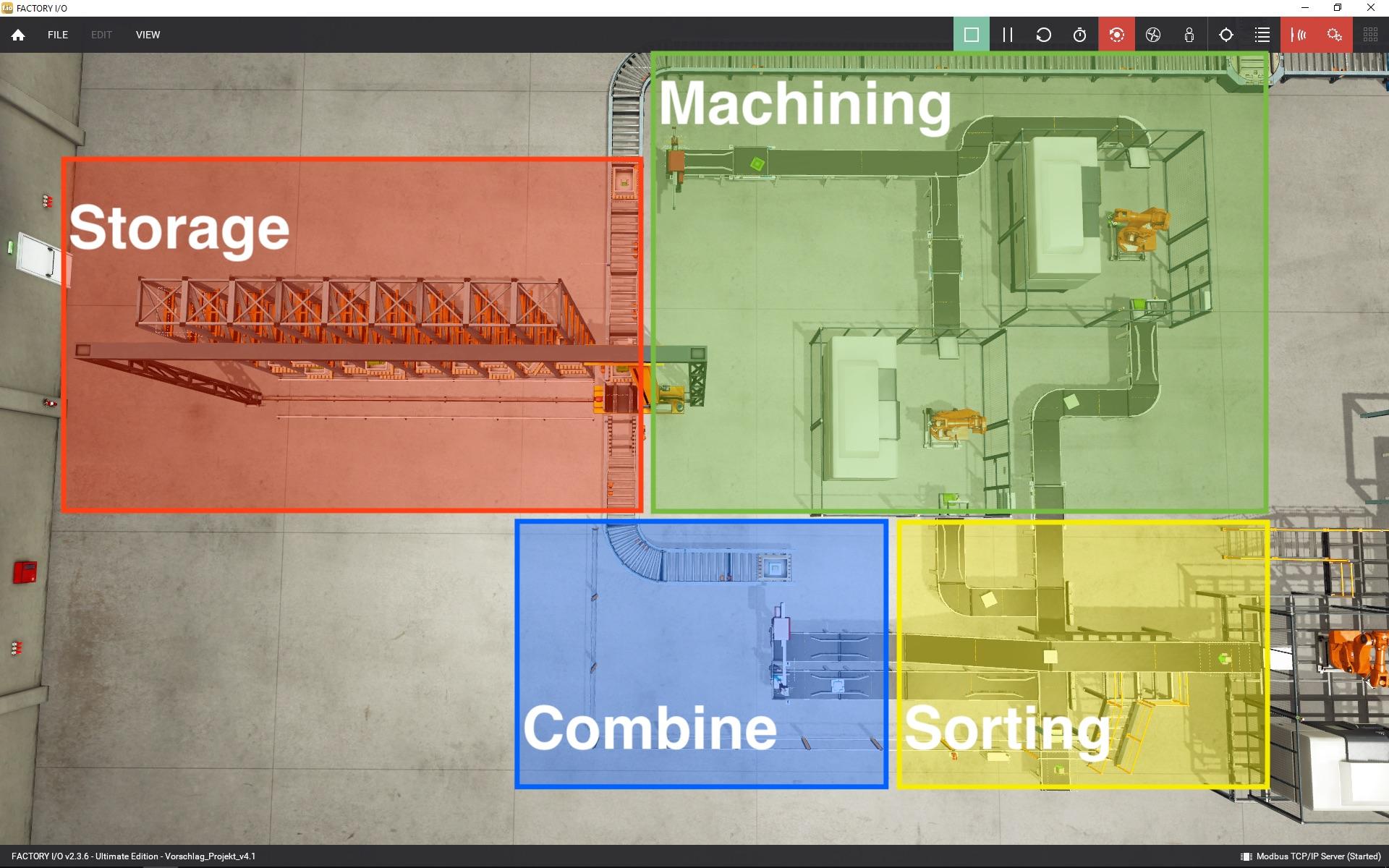}
    \caption{Overview across the four production cells in the \fio{} scenario\label{fig:factoryio-production-cells}}
\end{figure}

As mentioned above the virtual factory to be secured was built with \fio. The virtual factory, or the scenario as it is called in \fio, was explicitly designed to include mostly independent production cells that are each controlled by a single PLC (either OpenPLC or hardware). On the one hand, this is close to the organization of real production lines as it keeps each PLC program simpler. On the other hand, this compartmentalization allows to add more distinct PLCs to the network to assess more complex attacks. The communication between the PLCs and \fio\ is performed via the \mb\ protocol as it is a common denominator between \opc\ and the Siemens S7. As \fio\ only supports a single client we bridged the PLCs via a \mb\ server, that runs on one RaspberryPI. \autoref{fig:factoryio-production-cells} shows the four production cells that we designed and combined in \fio.


These four production cells cover some common use-cases in real factories while also including some components that can be visibly manipulated in an attack, e.g., a Pick-and-Place crane in the \emph{Combine} cell and the conveyor belts connecting all cells. These component were included due to our testbed's focus on demonstrability.

\subsection{Attack Simulation\label{subsec:attack_simulation}}
We designed and implemented three kinds of attacks on the OT testbed. All attacks assume the attacker is already inside the target network, has network access, and can do a scan of the subnet to find the target devices. While external attacks could also be considered through a firewall and separate subnet or VLAN, this is not the focus of the testbed.

\begin{itemize}
    \item Denial of Service (DoS) via a TCP-SYN flood on a PLC
    \item Sabotage via a forged \mb\ packet to manipulate internal state of production hardware
    \item Man-in-the-middle (MitM) attack to eavesdrop on factory communication and eventually transparently take over the whole factory
\end{itemize}

Before implementing these attacks, we performed a small evaluation of how attackers usually first gather intel about the target systems by scanning the subnet. As \mb\ is transmitted over TCP/IP, \emph{nmap}\footnote{\url{https://nmap.org}} can be used to detect open ports in a subnet. In our tests we could reliably verify this before implementing the attacks.

One of the most common forms of attack is the DoS. In a production setup the overload and subsequent shutdown of a single machine can in worst case stop the complete production process and is thus an easy way for an attacker to cause great financial damage to an organization. We performed a simple DoS by flooding a target PLC with TCP-SYN packets via \lstinline{hping} (shown in~\autoref{lst:hping}).

\begin{lstlisting}[language=sh, caption={Example command for a simple TCP-SYN flood DoS via \lstinline{hping}}, label={lst:hping}]
$ hping3 -S -d 120 -p 502 --flood <target-ip>
\end{lstlisting}

The arguments for \lstinline{hping} set the TCP-SYN flag (\lstinline{-S}), set the destination port to the common \mb\ port 502 (\lstinline{-p 502}) and enable flooding (\lstinline{--flood}).
The results on the virtual factory were as expected. The production cell that was controlled by the targetted PLC was completely stopped as the PLC was unable to reply to the \mb\ requests of \fio. While simple, this attack already shows that internal attackers can cause big damage with minimal effort. However a sophisticated attacker might not chose this attack in a real-world scenario as it can be easily detected on the network side, if security monitoring is in place.

Furthermore, we implemented a sabotage attack via the forgery of \mb\ packets. The basic idea is to send a specially crafted \mb\ packet and influence the production hardware (usually \mb\ clients) by corrupting their internal state. Of course, this attack is only possible as \mb\ offers no built-in authentication or other protection against unauthorized command execution.
One example for this is the Pick-and-Place crane in the \emph{Combine} production cell (see:~\autoref{subsec:setup}). To achieve this, we used the \lstinline{auxiliary/scanner/scada/modbusclient} module from \lstinline{Metasploit}. \autoref{lst:msf-forgery} shows an example script to attack the Pick-and-Place crane in this fashion.

\begin{lstlisting}[showstringspaces=false, language=sh, caption={\mb{}-based sabotage attack via \lstinline{msfconsole}}, label={lst:msf-forgery}]
$ msfconsole -x
"use auxiliary/scanner/scada/modbusclient; \
    set RHOST 192.168.1.62; \
    set RPORT 502; \
    set UNIT_NUMBER 0; \
    set ACTION WRITE_COILS; \
    set DATA_ADDRESS 34; \
    set DATA_COILS 1; \
    run;"
\end{lstlisting}

This attack demonstrates how an intruder can stealthily manipulate coils in a \mb\ client. Depending on the exact coil this can have big consequences as machines, e.g., degrade more quickly in a modified environment leading to an unexpected production stop when the machine has to be replaced or repaired. As \fio\ is somewhat limited in which machines it can simulate the attack on the Pick-and-Place crane in our scenario is easily detected in the factory. The attack misplaces the arm of the crane by 90 degrees and as a result the whole cell is blocked.

Finally, we also designed a Man-in-the-Middle attack via \lstinline{ettercap}. First, \lstinline{ettercap} uses ARP cache-poisoning to route traffic over itself. This allows the attacker to eavesdrop on all traffic in the subnet as \mb\ in its default configuration is not encrypted. Afterwards \lstinline{ettercap}'s filter function can be used to manipulate traffic at will. \autoref{lst:ettercap-filter1} shows one such filter that reverses the direction of some conveyor belts in the virtual factory.

\begin{lstlisting}[language=c, caption={\lstinline{ettercap} filter to reverse direction of conveyor belts}, label={lst:ettercap-filter1}]
if (ip.proto == TCP && tcp.dst == 502) {
  if (search(DATA.data, "\xff\xce")) {
    replace("\x00\x00\x00\xfa",
      "\xfb\x1d\xfb\x1d");
    replace("\x00\x32\x00\x00\x00\x00",
      "\xfb\x1d\xfb\x1d\xfb\x1d");
    replace("\x00\xfa", "\xfb\x1d");
    replace("\x00\x32", "\xfb\x1d");
    replace("\xff\xce", "\x01\xfa");
  }
}
\end{lstlisting}

First, the filter checks that the inspected traffic is TCP and on destination port \texttt{502} (the default \mb\ port). Then the payload of the packet is matched against a specific byte pattern \texttt{0xFFCE}. This pattern are \fio\ specific and identify the \mb\ message that sets the speed of the conveyor belts. We identified this pattern by monitoring the traffic similar to what an attacker would do. If both conditions are met, the filter transparently replaces the bytes responsible for controlling the speed of the conveyor belts to \texttt{0xFB1D} (equals \texttt{-1521} in two's complement) effectively reversing their direction. The multiple \lstinline$replace$ calls are required to catch all possible states the conveyor belts can be in.

This attack highlights how an attacker can eavesdrop on all communication inside a subnet and latter effectively rewrite all traffic relayed through their machine. The belt filter massively impacts the production process and results in work materials flying through the factory. While this was only an example, it clearly highlights that a Man-in-the-Middle attacker can control the whole factory and manipulate all devices (given they know the correct \mb\ coils and values).

\subsection{Security Monitoring\label{subsec:security_monitoring}}
To monitor the OT network we developed an anomaly detection component that ingests \mb\ traffic from the network via the mirror port of the switch. As our main goal was to evaluate new monitoring solutions we did not include a firewall or any other active component. All network data is buffered to disk by tShark and then processed in python programs. For analysis we considered two approaches, namely statistical anomaly detection and machine learning. Statistical anomaly detection proved to be difficult to apply as it valid changes in the communication patterns are not easy to distinguish from attacks. Therefore, we will detail the machine learning approaches in the following.

\begin{figure}\centering
    \includegraphics[width=.9\linewidth]{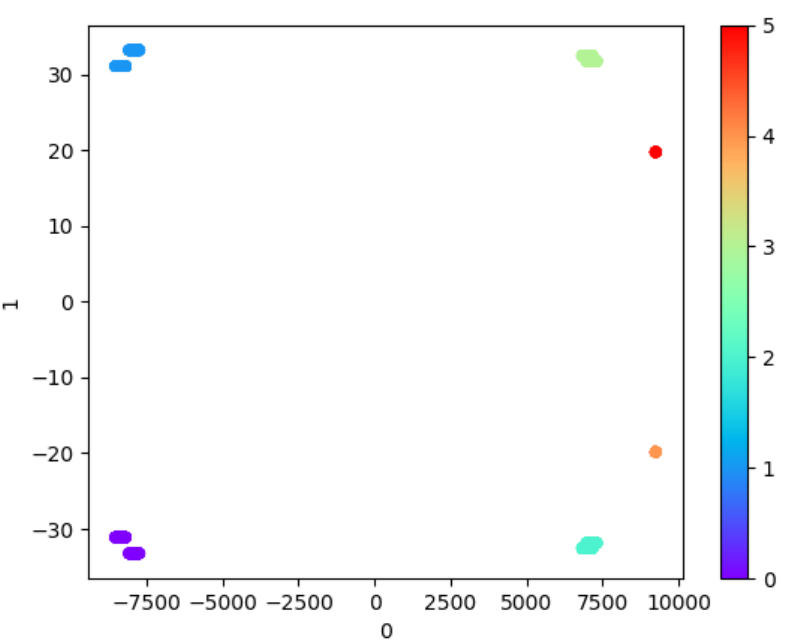}
    \caption{Example results for k-means with a subset of attributs and two outliers indicating attacks.}
    \label{fig:k-means}
\end{figure}

By parsing the network data we get several attributes from IP and TCP headers and the timing of the packets. We applied the k-means algorithm to cluster the data by frame number, time, time interval from previous packet, TCP sequence number, TCP acknowledgement number, TCP window size, source IP-address, source port, destination IP-address, and destination port. The results show that it is easy to cluster the data as a lot of values are repeated quite often. However, it is not that easy to find outliers resulting from attacks. By testing different subsets of attributes, we found that the combination of frame length, source port, destination port, and transaction identifier (modbus) gives far better results and attacks are clearly visible as outliers as Figure~\ref{fig:k-means} shows.

We also experimented with a neural network, namely a recurrent neural network (RNN) that allows to process normal attributes and timing information. We encountered some problems due to the amount of data required for training and the high configuration and tuning effort. Nethertheless, the RNN is able to correctly classify about 85\% of the traces as either normal or anomaly and other algorithms might perform better. 

\section{Conclusion\label{sec:conclusion}}
In this paper we showed a concept for a OT testbed to evaluate cyber attacks, possible counter measures and mitigations. By using commodity hardware in combination with a virtualized factory through \fio, the testbed remains portable and cost-effective, but can be combined with real automation hardware if desired.
We implemented a variant of this concept with RaspberryPis and a Siemens PLC with a focus on ease of demonstration.

Our implementation proved to be very cost-effective and highlights three important ways an attacker can influence the production process. \fio\ is great for visualizing malicious impact in a factory, as flying materials and missguided crane arms are clearly recognizable as problems. To be able to connect multiple hetergenous PLCs with \fio\ the \mb{}-based approach proved useful.
We plan to extend this work by performing more attacks and implementing better monitoring and attack-detection approaches.

\section*{Acknowledgment}
The authors would like to thank the students, Julian Brott, Keller, Elaha Khaleqi, Anton Laugwitz, Jule Lüderitz and Hanna Schambach, who helped in building and evaluating our testbed as well as Airbus Cybersecurity GmbH for contributing hardware and knowledge.

\bibliographystyle{IEEEtranN} 
\bibliography{komma2019}

\end{document}